\documentclass[RNASS]{aastex631}

\def\lessim{\mathrel{\hbox{\rlap{\hbox{\lower4pt\hbox{$\sim$}}}\hbox{$<$}}}}
\def\grtsim{\mathrel{\hbox{\rlap{\hbox{\lower4pt\hbox{$\sim$}}}\hbox{$>$}}}}

\def\equsim{\mathrel{\hbox{\rlap{\hbox{\raise3pt\hbox{$\sim$}}}\hbox{$=$}}}}

\shorttitle{V1674 Her}
\shortauthors{Quimby, Shafter \& Corbett}

\graphicspath{{./}{figures/}}

\begin{document}

\title{The Detailed Light Curve Evolution of V1674 Her (Nova Her 2021)}

\correspondingauthor{R. M. Quimby}
\email{rquimby@sdsu.edu}

\author[0000-0001-9171-5236]{R. M. Quimby}
\affiliation{Department of Astronomy and Mount Laguna Observatory\\
San Diego State University, San Diego, CA 92182, USA}
\affiliation{Kavli Institute for the Physics and Mathematics of the Universe (WPI)\\ The University of Tokyo Institutes for Advanced Study\\
The University of Tokyo, Kashiwa, Chiba 277-8583, Japan
}

\author[0000-0002-1276-1486]{A. W. Shafter}
\affiliation{Department of Astronomy and Mount Laguna Observatory\\
San Diego State University, San Diego, CA 92182, USA}

\author[0000-0002-6339-6706]{H. Corbett}
\affiliation{Department of Physics and Astronomy\\
University of North Carolina at Chapel Hill, Chapel Hill, NC 27599-3255, USA}

\begin{abstract}
We report high-cadence photometry of the ultra-fast ($t_2\sim1.2$~d)
nova V1674 Her during its rise to maximum light ($V\sim6.3$) and
the beginning of its subsequent decline.
These observations from Evryscope and the Mount Laguna Observatory
All-Sky Camera reveal a plateau in the
pre-maximum light curve at $g\sim14$ ($\sim$8 mag below peak) that
lasted for at least three hours. Similar features (so-called pre-maximum halts) have been
observed in some novae near maximum light, but to our knowledge the
detection of a plateau in the light curve $\sim$8 mag below peak is unprecedented.
\end{abstract}


\section{Introduction} \label{sec:intro}

V1674 Her (Nova Her 2021) was discovered by Seiji Ueda
(Kushiro, Hokkaido, Japan) on 2021 June 12.5484 UT
near magnitude 8.4.
The object subsequently rose to reach
a peak visual brightness of $V\equsim6$ on 2021 June 12.9 UT according to
observations reported to the AAVSO. The nova then faded rapidly,
declining by $\sim$2 mags in approximately a day. Novae with $t_2$ times
this short are extremely rare, and the object has been
monitored at multiple wavelengths
since its discovery
\citep[e.g.,][]{pag21,woo21,sok21,ayi21,mun21}.
The quiescent counterpart has been identified
and a recent analysis has revealed an 8.4-min period in the
progenitor binary that likely reflects the rotation period of the
accreting white dwarf \citep{mro21}.
In addition, recent spectroscopic
observations by \citet{wag21} have revealed strong
[Ne V] 342.6 nm and the [Ne III] pair at 386.9, 396.8 nm, which
establish the object as a member of the class of ``Neon Novae" \citep{geh98}.

In this {\it Research Note\/} we report photometric observations
of V1674 Her obtained with unprecedented coverage and temporal resolution
using the Evryscope and the All-Sky Camera (ASC)
at the Mount Laguna Observatory (MLO).
The Evryscope consists of
an array of 20 6.1-cm aperture telescopes that monitor $\sim7400$ square degrees 
simultaneously and continuously
every two minutes \citep{law14}. There are two Evryscopes currently in
operation: Evryscope South at Cerro Tololo, Chile (CTIO),
and Evryscope North at MLO. With a limiting magnitude reaching
$g\sim14$, the Evryscope is a useful instrument to monitor bright
transients such as classical novae. In the case of V1674 Her, which just
reached naked-eye brightness, the object proved to be too bright
for the Evryscope, with all measurements with $g\lessim9$
being saturated. Fortunately, MLO is equipped with an ASC that is nominally used
to monitor the weather during remote observing runs. The MLO-ASC 
is an Alcor-System OMEA\,8M that
records FITS images with 10~s integrations followed by $\sim$5~s for readout and processing. The ASC detector was cooled to $-3^{\circ}$~C, but dark current subtraction and flat-fielding have not yet been implemented. However, by stacking raw ASC images into roughly 5-min bins ($\sim20$ frames), we can
mitigate these signals
and reach limiting magnitudes of $\sim9$\,mag calibrated to the Gaia band.
Thus, the ASC nicely complements our Evryscope observations.

\begin{figure*}
\plotone{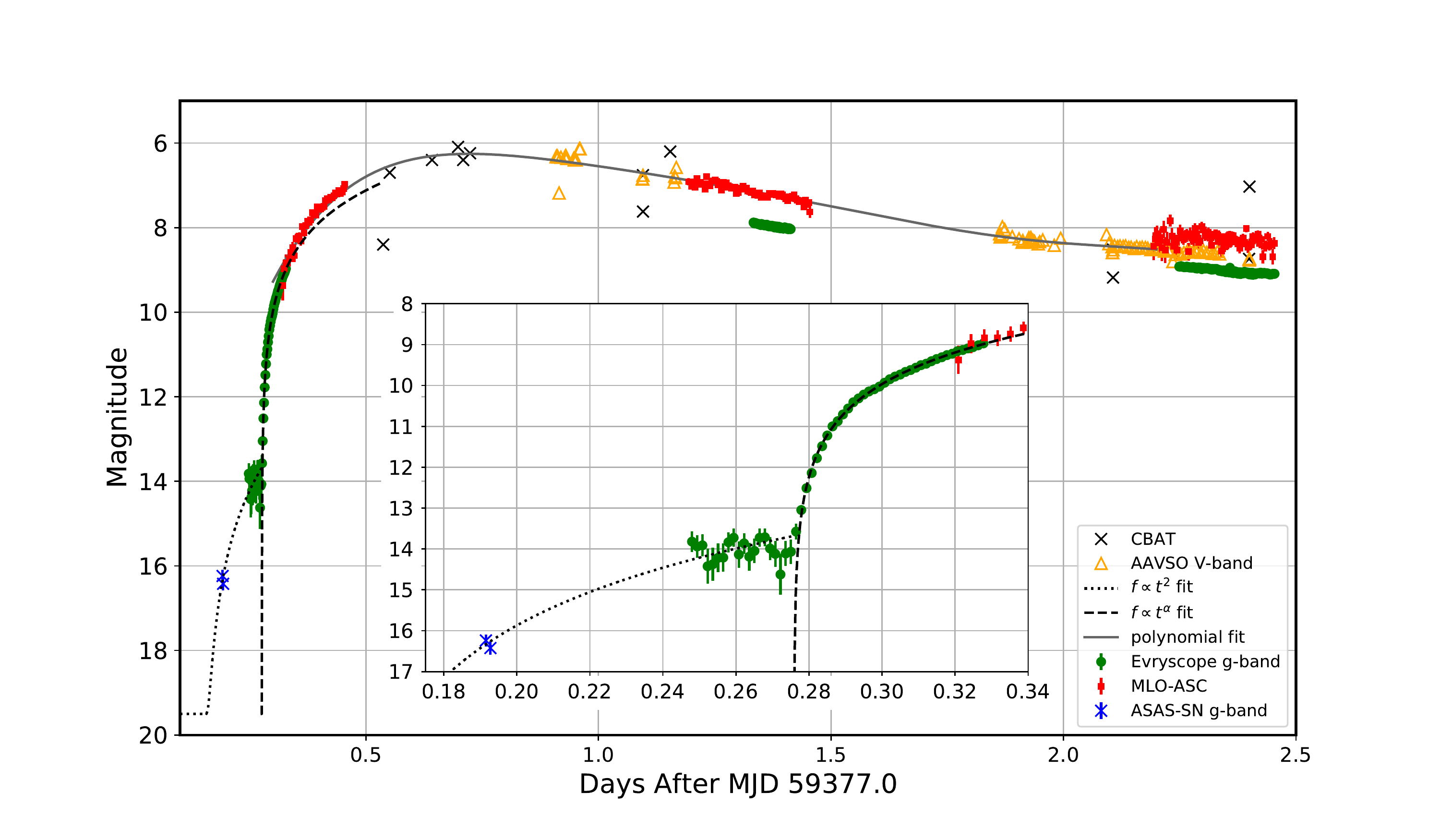}
\caption{The complete light curve of V1674 Her based primarily on our preliminary
Evryscope and ASC observations. The data have been augmented by photometric
measurements from the AAVSO, CBAT and ASAS-SN. The Evryscope observations
reveal pre-maximum plateau, lasting for $\sim$3 hr, that is highlighted
in the insert. Details of the various symbols and fits to the light
curve are given in the figure legend.}
\label{fig:lc}
\end{figure*}

\section{The Light Curve} \label{sec:lc}

Figure~\ref{fig:lc} shows our combined Evryscope and ASC light curve for V1674 Her.
The most striking feature is the plateau seen near $g=14$ in the pre-maximum
light curve. The aperture photometry may include some light from neighboring objects, but no source has been significantly detected at this position in 150 previous nights of Evryscope observations and visual inspection of the images suggest the appearance of a faint source between MJDs 59377.248 and 59377.275. The existence of the plateau is further supported by two photometric points from the All-Sky Automated Survey
for Supernovae \citep[ASAS-SN,][]{sha14} database --
the first on MJD 59377.1956 at $g=16.2$,
and the second on MJD 59377.1968 at $g=16.4$ --
that together with our Evryscope observations constrain the length
of the pre-maximum plateau to be less than 3~hr.
We are unaware of any additional photometry that would help constrain the
evolution of the nova prior to reaching the plateau.

We have fit the
early rise with a model of the early fireball as a uniformly expanding
spherical blackbody of radius $R(t)$ and (assumed constant)
effective temperature, $T_{\rm eff}$. 
The time evolution of the observed flux in such a model is given by:

\begin{equation}
f(t)=[\theta(t)]^2~\sigma~T_{\rm eff}^4,
\end{equation}
\noindent
where $\theta(t)=R(t)/d$ is the angular radius of the expanding photosphere.

Assuming that the fireball expands with constant velocity
(i.e., $R\propto~t$), we see that $f\propto~t^2$.
This simple model, which is shown as the dotted line in Figure~\ref{fig:lc}, allows us to estimate the time of the initial
explosion to be at approximately MJD 59377.1575. We note that this time is $\sim$3~hr earlier
than one would estimate based on an extrapolation of the later,
post-plateau, rise recorded by Evryscope between MJDs  59377.2793 and 59377.3278.  The best-fit power-law over this range is $f\propto~t^{1.2}$ (dashed line in Fig.~\ref{fig:lc}).

To estimate the time and magnitude at peak we have fit a 6th-order polynomial
to a combination of the ASC time-series and the AAVSO and CBAT data points.
The fit shows that peak brightness reaches $m\sim6.3$ (pseudo $V$ magnitude)
at MJD 59377.722,
and fades by two magnitudes in 1.18~d. This value is comparable with
other estimates of $t_2$ \citep[e.g.,][$t_2\lessim1$~d]{ayi21}, and suggests that V1674 Her is arguably
the fastest nova on record [the recurrent nova U~Sco also has a
$t_2$ time reported to be $\sim$1.2~d \citep{sch10}].

\section{Concluding Remarks}

V1674 Her is the first nova where the pre-maximum rise has been observed
in great detail thanks to the availability of modern automated sky
patrols (e.g., Evryscope and ASAS-SN). Observations made possible by these
instruments have allowed us to discover a pre-maximum plateau in the
light curve of V1674 Her well below maximum light. The plateau appears to exhibit fluctuations
of order 15~min that may be related to the 8.4~m periodicity reported by \citet{mro21}.
The plateau is reminiscent
of the pre-maximum halts observed just below peak in some novae \citep[e.g.,][]{hou10, hou16}, although
in V1674 Her the feature occurs relatively early in the eruption and well below maximum light.
The physical process or processes causing the plateau is
currently unclear, but may be related to radiation from a precursor UV flash or the result of
changes in the structure of the convective zone in the nova's expanding photosphere
\citep[e.g., see][]{hil14}.
Further modeling will
be required to fully understand the nature of the pre-maximum plateau
in V1674 Her.

\bibliography{novaher21}{}
\bibliographystyle{aasjournal}

\end{document}